\documentclass[twocolumn,showpacs,aps,prl,superscriptaddress]{revtex4}
\usepackage{graphicx}
\usepackage{dcolumn}
\usepackage{amsmath}
\usepackage{epsfig}
\usepackage{colordvi}
\usepackage{color}
\usepackage{array, longtable}
\usepackage{hhline}

\newcommand{\BaBarYear}{2006}
\newcommand{\BaBarNumber}{1482}
\newcommand{\SLACPubNumber}{12146}

 \newcommand{\BaBarType}      {PUB}  

\input pubboard/babarsym

\def\KKppch    {\ensuremath {K^+K^-\pip\pim}\xspace}
\def\KKppnt    {\ensuremath {K^+K^-\pi^0\pi^0}\xspace}
\def\mKKpp     {\ensuremath {m(\Kp\Km\pi\pi)}\xspace}
\def\chifourpi {\ensuremath {\chi^2_{4\pi}}\xspace}
\def\chiKKppch {\ensuremath {\chi^2_{KK\pip\pim}}\xspace}
\def\chiKKppnt {\ensuremath {\chi^2_{KK\pi^0\pi^0}}\xspace}

\long\def\inst#1{\par\nobreak\kern 4pt\nobreak
    {\it #1}\par\vskip 10pt plus 3pt minus 3pt}

\begin{document}

%


\begin{flushleft}
\babar-\BaBarType-\BaBarYear/\BaBarNumber \\
SLAC-PUB-\SLACPubNumber
\end{flushleft}


\title{\large \bf
\boldmath

A Structure at 2175~\mev in $\epem \!\!\to\! \phi f_0(980)$ Observed
via Initial-State Radiation
} 

%
\author{B.~Aubert}
\author{M.~Bona}
\author{D.~Boutigny}
\author{F.~Couderc}
\author{Y.~Karyotakis}
\author{J.~P.~Lees}
\author{V.~Poireau}
\author{V.~Tisserand}
\author{A.~Zghiche}
\affiliation{Laboratoire de Physique des Particules, IN2P3/CNRS et Universit\'e de Savoie,
 F-74941 Annecy-Le-Vieux, France }
\author{E.~Grauges}
\affiliation{Universitat de Barcelona, Facultat de Fisica, Departament ECM, E-08028 Barcelona, Spain }
% ...
\collaboration{The \babar\ Collaboration}
\noaffiliation

\date{\today}


\begin{abstract}
We study the initial-state-radiation processes 
$\epem \!\!\to\! K^+ K^- \pipi\gamma$ and
$\epem \!\!\to\! K^+ K^-  \ppz\gamma$
using an integrated luminosity of
232~\invfb collected at the $\Upsilon(4S)$ mass with the \babar\ detector at SLAC. 
Even though these reactions are dominated by intermediate states with excited
kaons, we are able
to study for the first time the cross section for  
$\epem \!\!\to\! \phi(1020) f_{0}(980)$ as a function of center-of-mass energy.
We observe a structure near threshold consistent with a 
$1^{--}$ resonance with mass 
$m\! =\! 2.175 \pm 0.010\pm 0.015~\gevcc$ and width 
$\Gamma\! =\! 58\pm 16\pm 20~\mev$. 
We observe no $Y(4260)$ signal and set a limit of
$\BR_{Y\to\phi\pipi}\cdot\Gamma^{Y}_{ee}<0.4$~\ev (90\% confidence level), 
which excludes some models.
\end{abstract}

\maketitle


The nature of the $Y(4260)$ resonance, 
which \babar\ recently discovered~\cite{y4260} through its
production via initial state radiation (ISR) in \epem annihilations
and its decay into $J/\psi\pipi$, remains unclear.
It is well above threshold for the $D^{(*)}\overline{D}^{(*)}$ decays
expected for a wide charmonium state, but no peak is observed in the
total cross section $\epem\to hadrons$ in this mass region.
Some models~\cite{shin} predict a large branching fraction for
$Y(4260)$ into $\phi\pi\pi$.
Moreover, the rich spectroscopy of the $J/\psi\pi\pi$ final state
motivates a thorough investigation of the analogous $\phi\pi\pi$ state. 

In this paper we update our previous analysis with ISR of $\epem
\!\to K^+K^-\pipi$ ~\cite{isr4pi}. 
We include more data and relax the selection criteria, 
resulting in a fivefold increase in the number of selected events.
We obtain an improved $\epem \!\!\to\! K^+K^-\pipi$ cross section measurement
over a wide range of effective \epem center-of-mass (C.M.) energies,
and perform the first studies of the $\phi\pipi$, $f_0(980)K^+K^-$ and 
$\phi f_0$ intermediate states.
We also present the first measurements of the $\epem \!\!\to\! K^+K^-\ppz$ 
cross section and its $\phi f_0$ component.

We use data corresponding to an integrated luminosity of 232~\invfb
recorded by the \babar\ detector~\cite{babar} on and off the $\Upsilon(4S)$
resonance.
Charged-particle tracking is provided by a five-layer silicon vertex
tracker (SVT) and a 40-layer drift chamber (DCH) in
a 1.5\,T axial magnetic field. 
Photon and electron energies are measured in a CsI(Tl)
electromagnetic calorimeter (EMC).
Charged particles are identified by specific ionization 
in the SVT and DCH, and an internally reflecting ring-imaging
Cherenkov detector (DIRC). 

We use a simulation package developed for radiative processes that
generates hadronic final states following Ref.~\cite{kuehn2},
multiple soft photons from the initial-state using a structure-function 
technique~\cite{kuraev,strfun},
and photons from the final-state particles using PHOTOS~\cite{PHOTOS}.  
We generate $K^+K^-\pi\pi$ final states both according to phase space 
and with a model that includes the $\phi(1020) \!\to\! K^+K^-$ and 
$f_{0}(980) \!\to \pi\pi$ channels.
We pass the events through a detector simulation~\cite{GEANT4}, and
reconstruct them in the same way as we do the data. 
We generate a number of backgrounds with this package, 
including the ISR processes $\epem \!\!\to\! \pipi\pipi\gamma$, 
$\pipi\ppz\gamma$, $\phi\eta\gamma$, $\phi\pi^0\gamma$ and $\pipi\pi^0\gamma$, and we
also study $\epem \!\!\to\! q\qbar$ events generated by JETSET~\cite{jetset},
$\epem \!\!\to\! \tau^+\tau^-$ by KORALB~\cite{koralb},
and \Y4S decays using our own generator~\cite{evtgen}.

The initial selection of events with a high-energy photon
recoiling against a set of charged
particles and photons is described in Refs.~\cite{isr4pi,isr6pi}.
Here we accept all charged tracks 
that extrapolate to the interaction region,
and photon candidates with an EMC energy greater than 30~\mev.
The reconstructed vertex of the set of charged tracks
is used as the point of origin for all photons.

For each four-track event with one or two identified $K^\pm$, 
we perform a set of three-constraint kinematic fits
(see Ref.~\cite{isr6pi}).
We assume the photon with the highest C.M.\ energy to be from ISR,
and the fits use its direction, along with the four-momenta and covariance 
matrices of the initial \epem and the reconstructed tracks.
A fit using the $\pipi\pipi$ hypothesis returns a \chifourpi.
If the event contains an identified $K^+$ and $K^-$,
we fit to the $K^+K^- \pipi$ hypothesis and require $\chiKKppch \! <\!
30$. 
For events with one identified kaon, we perform fits with each of
the two oppositely charged tracks given the kaon hypothesis, and the
combination with the lowest \chiKKppch is retained if it is lower than
$30$ and $\chifourpi \! >\! \chiKKppch$.

For the events with two tracks, both identified as charged kaons,
and five or more photon candidates, 
all non-ISR photons are paired, and combinations lying within 35~\mevcc
of the $\pi^0$ mass are considered $\pi^0$ candidates.
We perform a six-constraint fit to each set of two non-overlapping 
$\pi^0$ candidates plus the ISR photon and the \Kp and \Km tracks,
and the combination with the lowest \chiKKppnt is retained if
$\chiKKppnt\! <\! 50$.
To suppress ISR $\Kp\Km\pi^0$ and $\Kp\Km \eta$ events,
in which photons from an energetic $\pi^0$ or $\eta$ 
combine with soft background clusters to form two $\pi^0$ candidates,
we reject events with large differences between the two photon energies 
in both $\pi^0$ candidates. 
The fitted three-momenta for each charged track and photon are used 
in further kinematical calculations.

We consider three types of backgrounds.
The first, 
which peaks at low values of \chisq, is 
due to non-ISR events, and is dominated by $\epem \!\!\to\! \qqbar$
events with a hard $\pi^0$ producing a fake ISR photon.
To evaluate this background, we use simulated mass and \chisq
distributions normalized to data events in which
the ISR photon combines with another cluster to form a $\pi^0$ candidate.
The second type of background, due to ISR
$\epem \!\!\to\! \pipi\pi\pi$ events with misidentified  
$\pi^\pm$, also contributes at low \chisq values.
We derive reliable estimates of their contributions from the known
cross sections~\cite{isr4pi}.
The third type of background comprises all remaining background
sources and is estimated from the control regions
$30\! <\!\chiKKppch\! <\!60$ and $50\! <\!\chiKKppnt\! <\!100$,
as detailed in Refs.~\cite{isr4pi,isr6pi}.
We subtract these backgrounds, about 8-10\% (15--20\%) total
contribution, from the selected $K^+K^-\pipi$ ($K^+K^-\ppz$) events.

We measure the track-finding efficiency from the data, 
and measure the kaon identification efficiency from a clean
sample of ISR $\epem \!\!\to\! \phi \!\!\to\! K^+K^-$ events
to a precision of 2.0\%,
a fourfold improvement over our previous result~\cite{isr4pi}.
The $\pi^0$ reconstruction efficiency is
determined from  ISR $\epem \!\!\to \omega\pi^0\gamma \!\!\to
\pipi\ppz\gamma$ events and the method described in Ref.~\cite{isr6pi}. 
The above procedures allow us to correct the efficiency obtained from the MC
simulation.
\begin{figure}[t]
\begin{center}
\includegraphics[width=0.9\linewidth, height=0.9\linewidth]{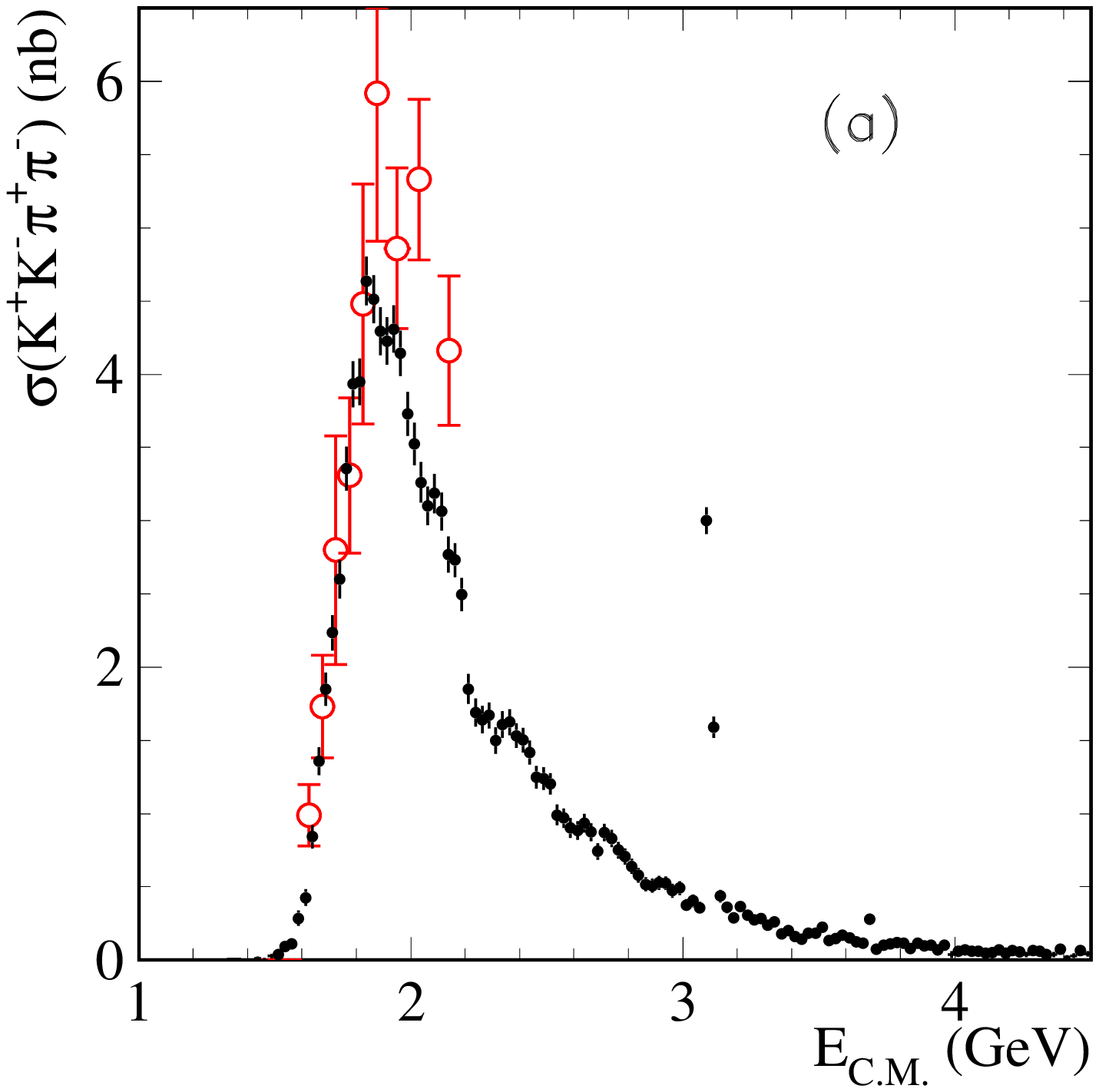}
\includegraphics[width=0.9\linewidth, height=0.9\linewidth]{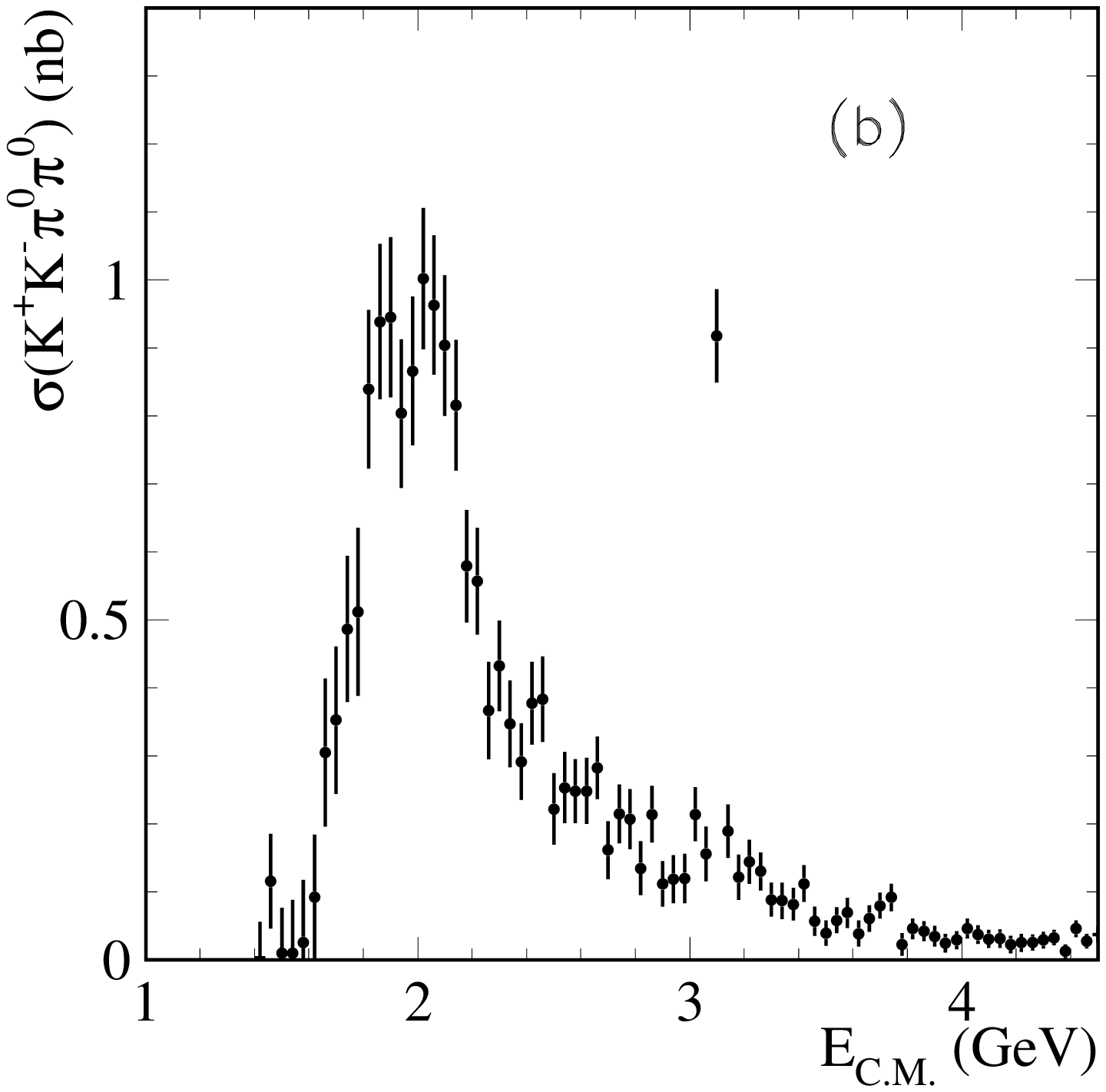}
\vspace{-0.4cm}
\caption{The
  a) $\epem \!\to K^+ K^-\pipi$ and 
  b) $\epem \!\to K^+ K^-\ppz$ cross sections as a function of \epem 
  C.M.\ energy.
  The direct measurements by DM1~\cite{2k2pidm1} are shown for comparison as
  open circles.
  Only statistical errors are shown.
  }
\label{2k2pi_ee_babar}
\vspace{-0.5cm}
\end{center}
\end{figure} 
\begin{figure}[t]
\begin{center}
\includegraphics[width=0.49\linewidth]{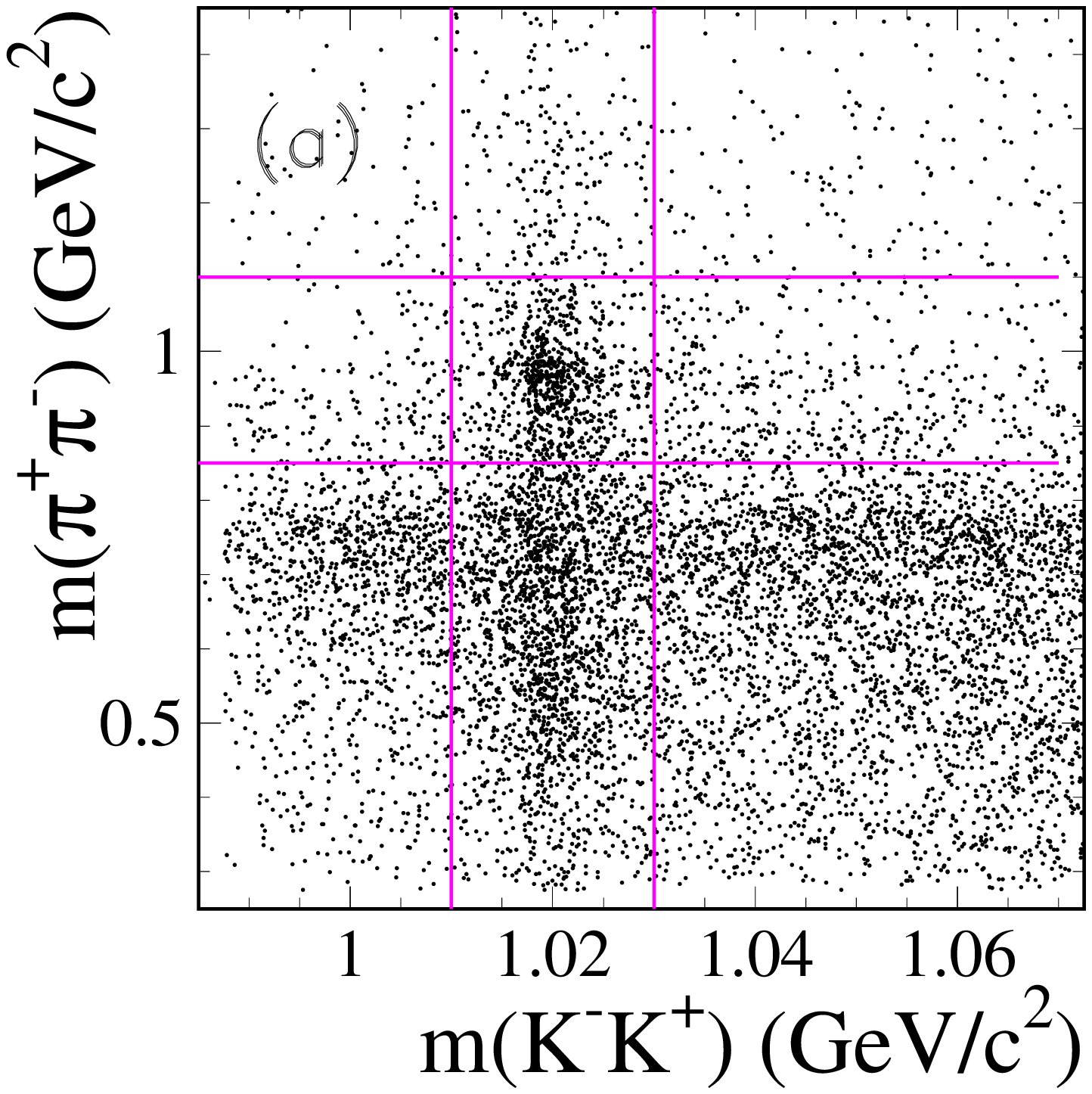}
\includegraphics[width=0.49\linewidth]{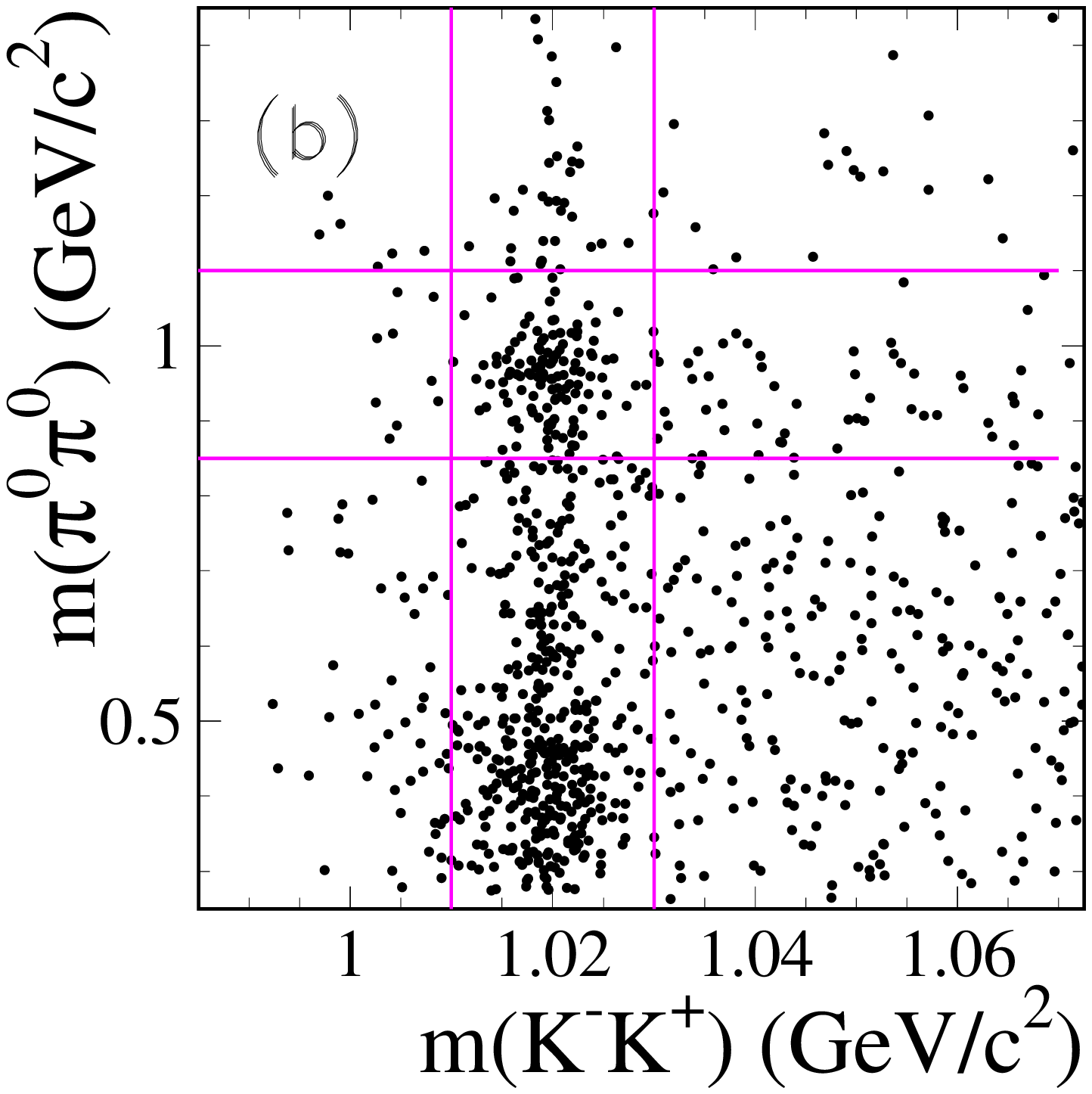}
\vspace{-0.7cm}
\caption{The
  scatter plots of the reconstructed a) $m(\pipi)$ and b)~$m(\ppz)$
  versus $m(K^+K^-)$ for selected events in the data.  
  The vertical (horizontal) lines bound a 
  $\phi$ ($f_0(980)$) signal region.
  }
\label{phif0_show1}
\vspace{-0.5cm}
\end{center}
\end{figure}
\begin{figure}[tbh]
\begin{center}
\includegraphics[width=\linewidth]{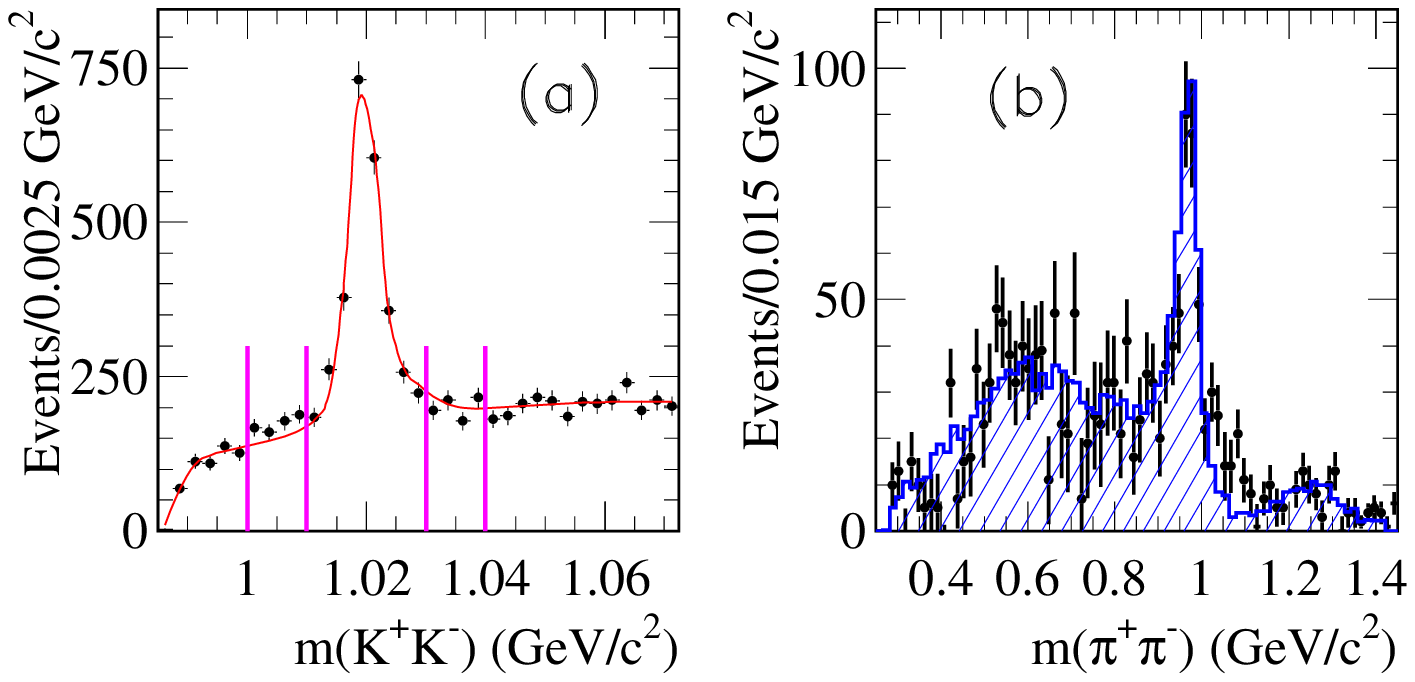}
\includegraphics[width=\linewidth]{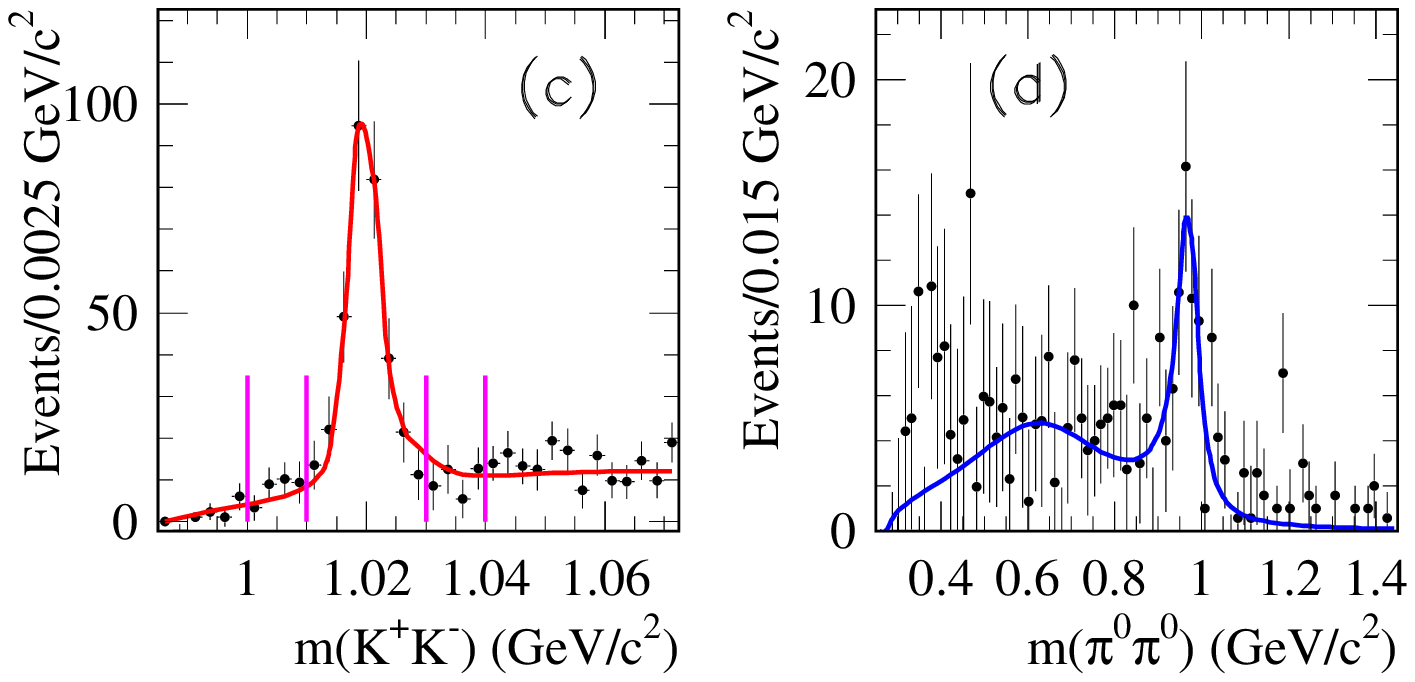}
\vspace{-0.7cm}
\caption{The
  $m(K^+K^-)$ projections for the a) \KKppch and c)~\KKppnt
  candidates in the data.
  The vertical lines delimit $\phi$ signal and sideband regions.
  b,d) $m(\pi\pi)$ distribution for events in the
  $\phi$ signal region of (a,c) minus that for events in the sidebands.
  The curves (histogram) represent the results of the fits (simulation)
  described in the text.
  }
\label{phif0_sel1}
\vspace{-0.5cm}
\end{center}
\end{figure}
In Fig.~\ref{2k2pi_ee_babar} we show the cross sections for the two
processes, calculated by dividing the
background-subtracted yield in each bin by the efficiency
and the ISR luminosity~\cite{isr4pi}.
The errors are statistical only.
The $\epem \!\!\to\! \Kp\Km\pipi$ cross section (Fig.~\ref{2k2pi_ee_babar}a)
is consistent with both the direct measurement by DM1~\cite{2k2pidm1}
and our previous measurement~\cite{isr4pi}, but is far more precise.
In addition to the sharp $J/\psi$ peak, wider structures are visible 
near 1.8~\gev, 2.2~\gev and possibly 2.4~\gev.
The $\epem \!\!\to\! \Kp\Km\ppz$ cross section (Fig.~\ref{2k2pi_ee_babar}b)
shows the same general features, including a $J/\psi$ peak and a
steep drop around 2.2~\gev.
The total systematic uncertainty in the $K^+K^-\pipi(\ppz)$ cross section
ranges from 7\% (10\%) at threshold to 9\% (15\%) at high $E_{C.M.}$.

As seen previously~\cite{isr4pi}, 
there is a rich substructure in the $\epem \!\!\to\! \KKppch$ process,
dominated by the $K^{*0}(892) K\pi$ intermediate state, 
but with large signals from the $K_1(1270)$, $K_2^{*0}(1430)$ and $K_1(1400)$ 
resonances.
The $\epem \!\!\to\! \KKppnt$ process is also dominated by
the $K^{*\pm}(892) K^{\mp}\pi^0$ intermediate state.
Understanding these contributions via a partial wave analysis
is outside the scope of this paper.

Here we concentrate on events with an intermediate $\phi(1020)$ and/or
$f_0(980)$ state.
Figure~\ref{phif0_show1} shows scatter plots of $m(\pipi)$ or
$m(\ppz)$ versus $m(K^+K^-)$ for the selected events (including
backgrounds) in the data.
A $\phi \!\!\to\! K^+K^-$ band is visible in both cases, 
as well as a concentration of events indicating
correlated production of $\phi$ and $f_0$.
A horizontal $\rho(770)$ band is visible for the charged mode only, 
and is due to $K_1 \!\to K\rho$ decays.
Most of the $K^*$ intermediate states are outside the bounds
of these plots.
Selecting $\phi$ events with
$|m(\Kp\Km)\! - 1020\mevcc|\! <\!10\mevcc$,
and subtracting events with 
$10\! <\! |m(\Kp\Km)\! - 1020\mevcc|\! <\!20\mevcc$ 
(see Figs.~\ref{phif0_sel1}a,c) and MC simulated backgrounds,
we obtain the $\phi$-associated $m(\pi\pi)$ distributions shown in 
Figs.~\ref{phif0_sel1}b,d.
Clear $f_0(980)$ signals are visible in both cases,  
and there is an indication of $f_2(1270) \!\to\! \pipi$. 
The histogram in Fig.~\ref{phif0_sel1}b is the result of a simulation
that includes $f_0(600)$, $f_0(980)$ and a small fraction of $f_2(1270)$
resonances and describes the 
general features of the distribution.
The curve in Fig.~\ref{phif0_sel1}d shows  a fit of two Breit-Wigner
functions corresponding to the $f_0(600)$ and $f_0(980)$ with the
relative phase set to $\pi$; events with $m(\ppz)<0.45$~\gevcc are dominated by
background-subtraction uncertainties and not used in the fit.
The fitted $f_0$ parameters are consistent with PDG~\cite{PDG} values.
\begin{figure}[t]
\begin{center}
\includegraphics[width=0.49\linewidth]{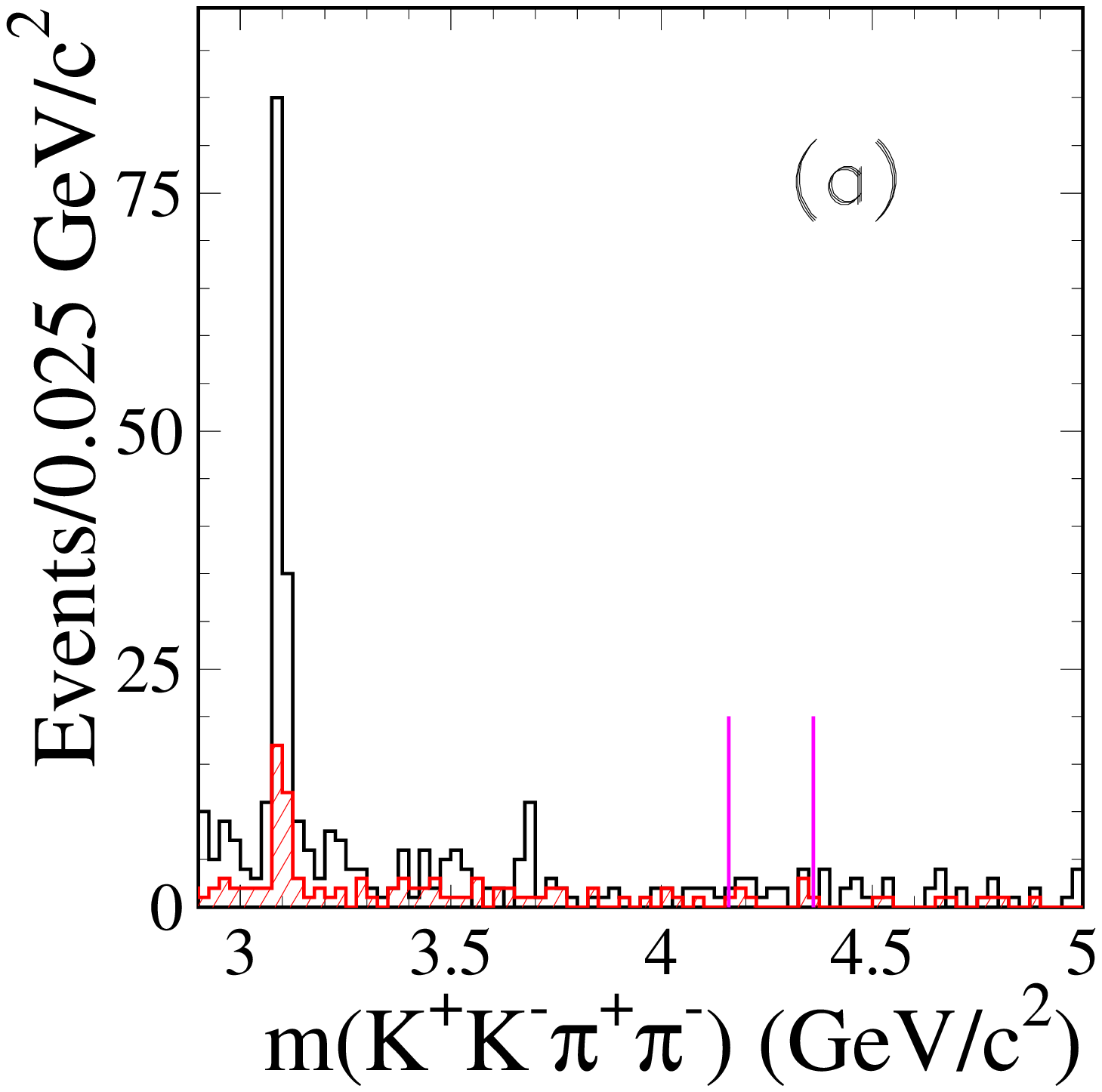}
\includegraphics[width=0.49\linewidth]{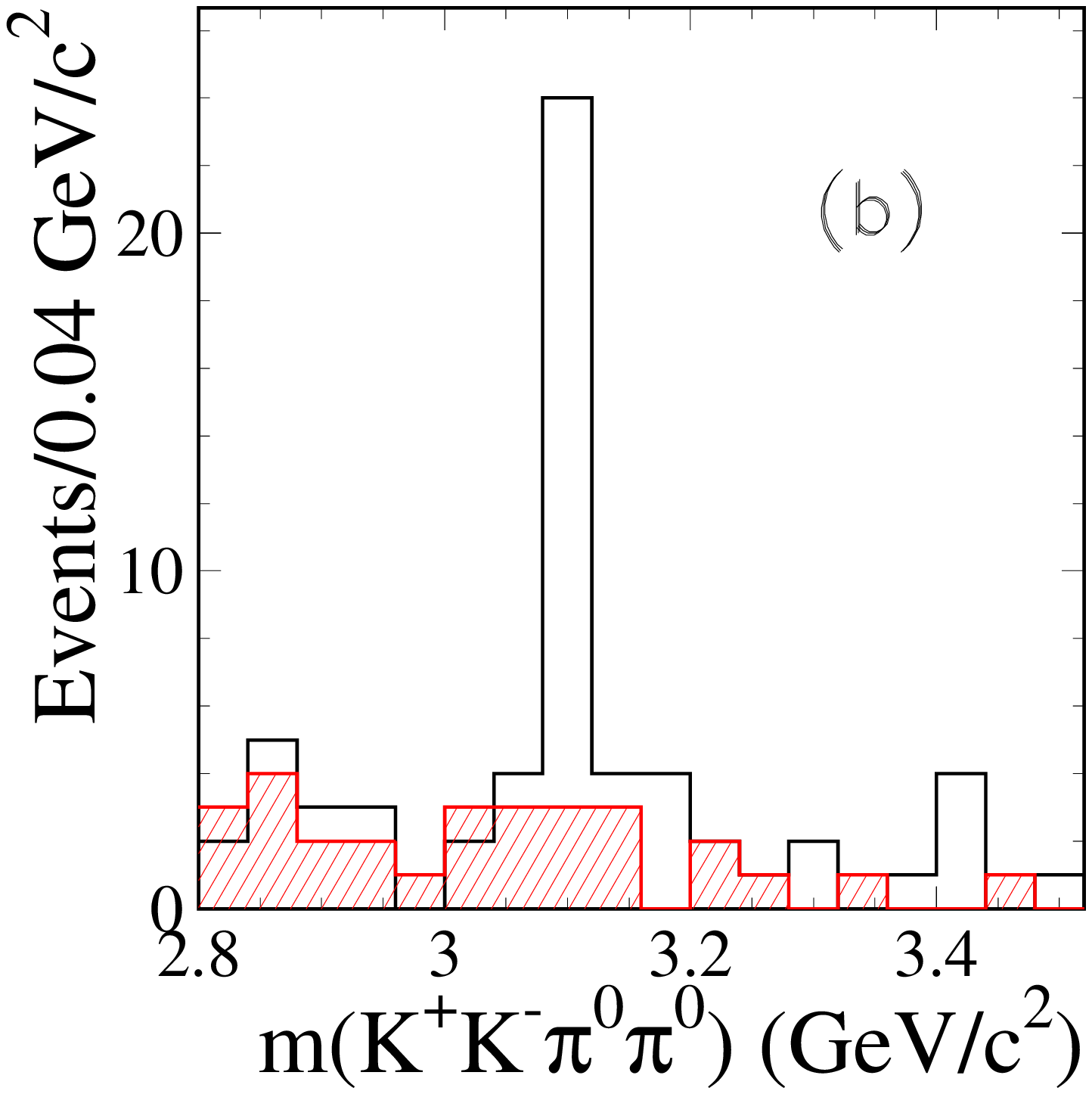}
\vspace{-0.7cm}
\caption{ The
  a) $m(\KKppch)$ and
  b) $m(\KKppnt)$ distributions in the
  charmonium region for events 
  in the $\phi$ signal (open histogram) and sideband (shaded
  histogram) regions.
  The vertical lines indicate the range used for the $Y(4260)$ 
  search.
}
\label{jpsi_phi2pi}
\vspace{-0.5cm}
\end{center}
\end{figure}
Figure~\ref{jpsi_phi2pi} shows the \mKKpp distributions in the
charmonium region for events with $m(\Kp\Km)$ in the $\phi$ signal 
and sideband regions.
There is a strong $J/\psi$ signal in both samples;
from the signal-sideband differences of 103$\pm$12 and 23$\pm$6 events,
we calculate\\
$\BR_{J/\psi\to\phi\pipi}\cdot\Gamma^{J/\psi}_{ee} \cdot 
 \BR_{\phi\to K^+K^-}  = (2.61\pm 0.30\pm 0.18)~\ev$\\
and the first measurement of\\
$\BR_{J/\psi\to\phi\ppz}\cdot\Gamma^{J/\psi}_{ee} \cdot \BR_{\phi\to
   K^+K^-}  =  (1.54\pm  0.40\pm 0.16)~\ev. $\\
We also observe 10$\pm$4 $\psi(2S) \!\to\! \phi\pipi$ decays,
from which we determine\\
$\BR_{\psi(2S)\to\phi\pipi}\cdot\Gamma^{\psi(2S)}_{ee}\cdot \BR_{\phi\to K^+
   K^-}=(0.28\pm 0.11\pm 0.02)~\ev.$

There is no signal for $Y(4260) \!\to\! \phi\pipi$.
In the region $|m(\phi\pipi)-m(Y)|\! <\! 0.1$~\gevcc
we find 10 events, and assuming a uniform distribution we estimate 9.2
background events from the 3.8--5.0~\gevcc region. 
This corresponds to upper limits of 5.0 events and
$$\BR_{Y \to \phi\pipi} \cdot \Gamma^{Y}_{ee}\! <\! 0.4~\ev$$
 at the 90\% confidence level, which 
is in agreement with the upper limit obtained by CLEO~\cite{ycleo} and
is well below our analogous measurement 
$\BR_{Y\to J/\psi\pipi} \!\cdot \Gamma^{Y}_{ee}\! =\!
 (5.5\pm 1.1^{+0.8}_{-0.7})~\ev$~\cite{y4260}. This excludes
models (e.g.~\cite{shin}) in which these two $Y(4260)$ branching
fractions are comparable.

\begin{figure}[t]
\includegraphics[width=0.9\linewidth, height=0.9\linewidth]{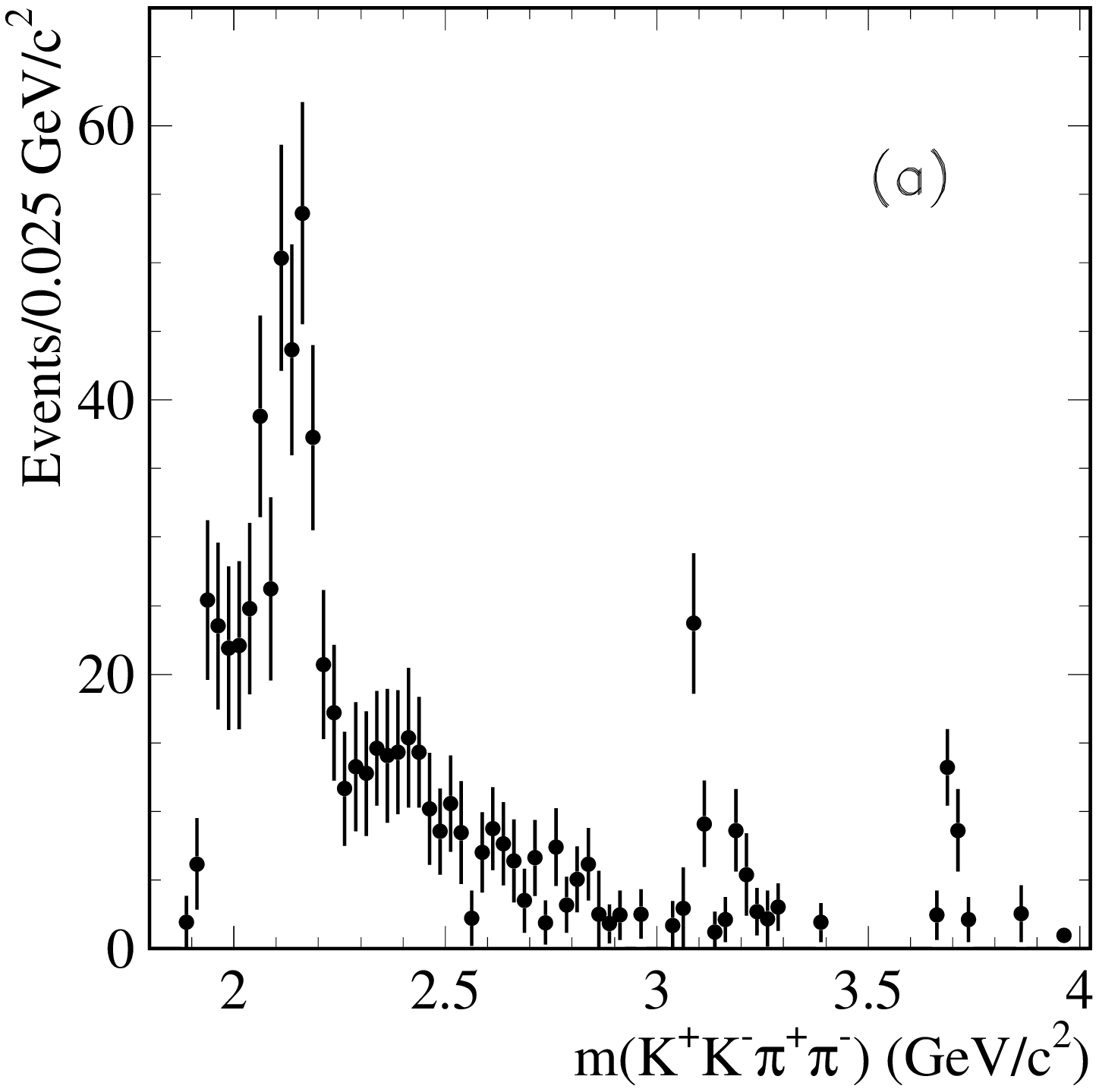}
\includegraphics[width=0.9\linewidth, height=0.9\linewidth]{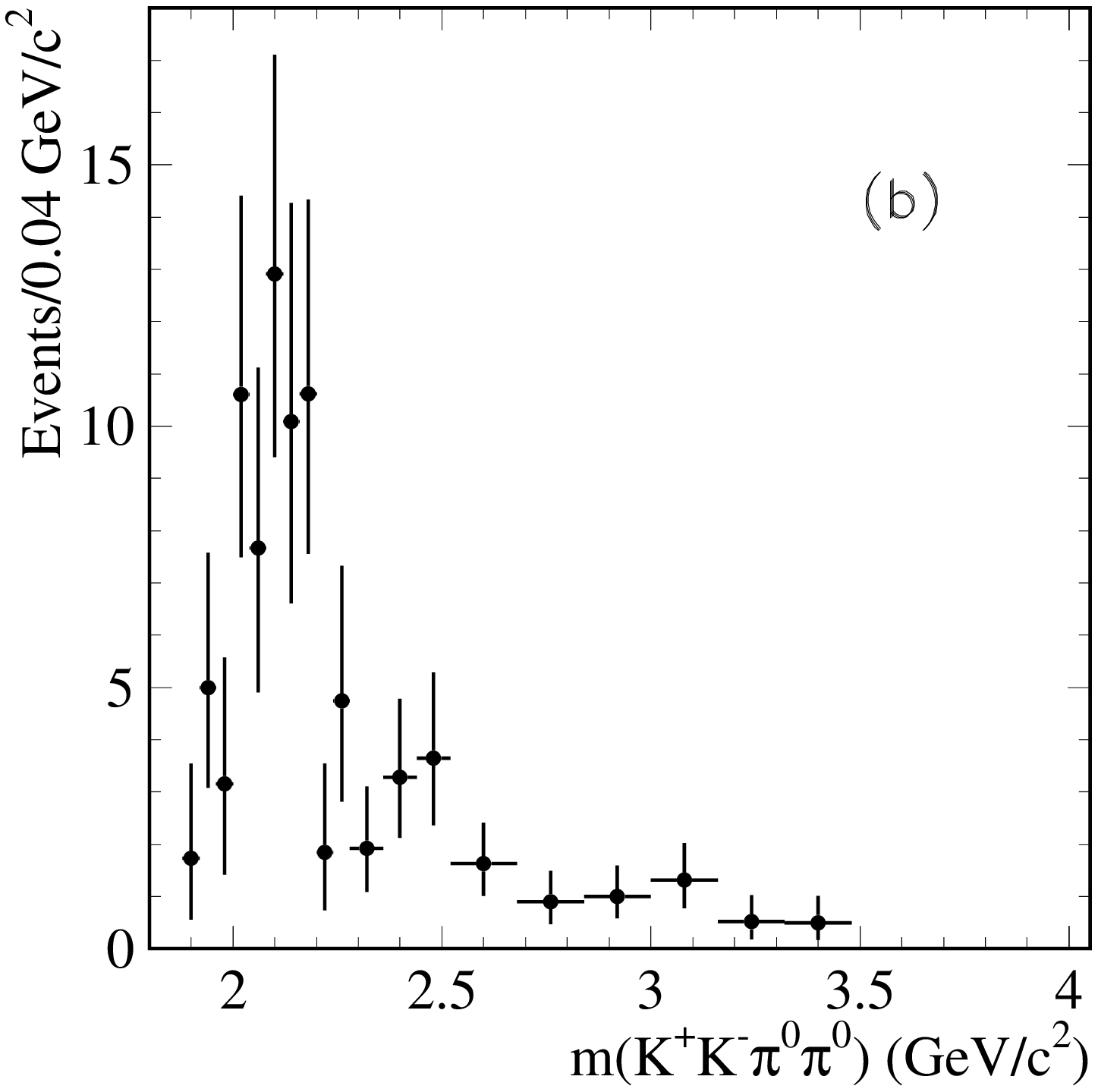}
\vspace{-0.4cm}
\caption{
  The number of a) $\epem \!\!\to\! \phi f_{0} \!\to\! K^+ K^-\pipi$ 
  and b)~$\epem \!\!\to\! \phi f_{0} \!\to\! K^+ K^-\ppz$ events
  vs.\ invariant mass
  extracted as described in the text.
  Some bins have been combined for clarity, as indicated by the
  horizontal error bars.
  }
\label{phif01}
\vspace{-0.3cm}
\end{figure}

We now consider the quasi-two-body intermediate
state $\phi f_{0}(980)$.
In each 25~\mevcc (40~\mevcc) bin of \mKKpp we select \KKppch
(\KKppnt) events with $m(\pipi)$ ($m(\ppz)$) in the 0.85--1.1~\gevcc region
and fit their $m(\Kp\Km)$ distribution 
to extract the number of events with a true $\phi$.
These are shown in
Fig.~\ref{phif01} with about 700 events for the $\Kp\Km\pipi$
channel and about 120 events for the $\Kp\Km\ppz$ channel; there is a
contribution of about 10\% from 
$\epem \!\to \phi\pi\pi$ events where the pion pair is not produced through
the $f_0(980)$.
Both distributions show the sharp rise from threshold as expected for a pair
of relatively narrow resonances, and a slow, smooth decrease at high
$E_{C.M.}$, with signals for $J/\psi$ and $\psi(2S)$ in Fig.~\ref{phif01}a.
Both also show a resonance-like structure at  about 2.15~\gevcc.
There are no known meson resonances with I=0 near this mass.
\begin{figure}[t]
\includegraphics[width=0.9\linewidth, height=0.9\linewidth]{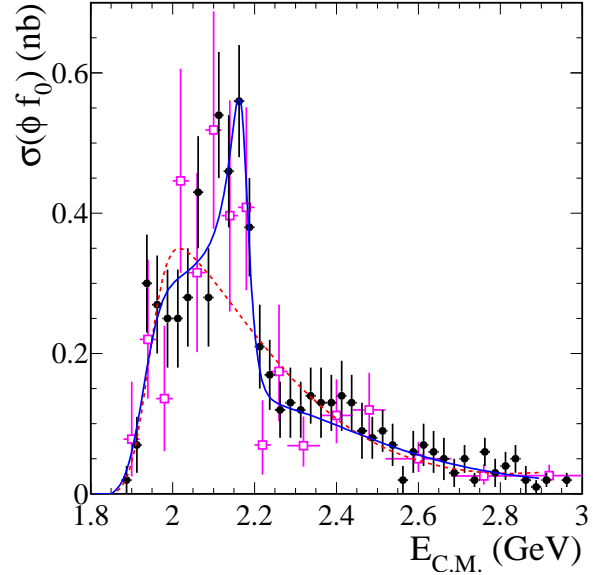}
\vspace{-0.4cm}
\caption{
  The $\epem \!\to\phi(1020) f_{0}(980)$ cross section, 
with about 10\% of the $\phi\pi\pi$ contribution,
obtained via  ISR in the $K^+ K^-\pipi$ (circles) 
  and $K^+ K^-\ppz$ (squares) final states.
  The curves represent results of the fits described in the text.
}
\label{phif0xsall}
\vspace{-0.3cm}
\end{figure}

Dividing by the efficiency, ISR luminosity, 
$\BR_{\phi\to K^+ K^-}\! =\! 0.491$~\cite{PDG}, and
$\BR_{f_0\to\pipi(\ppz)}\! =\! 2/3(1/3)$,
we obtain the two consistent measurements of the $\epem \!\to\! \phi f_0$ 
cross section shown in Fig.~\ref{phif0xsall} 
(including about 10\% $\phi\pi\pi$ contribution).
We use the following function of $s= E^2_{C.M.}$:
\begin{eqnarray}
\sigma(s)   &\!\! =\!\! & \frac{P(s)}{P(m_x^2)}  \cdot
    \left| A_{nr}e^{-i\psi_x} + \frac{\sqrt{\sigma_0} m_x \Gamma_x}
                 {m_x^2-s-i\sqrt{s}\Gamma_x} \right|^2 ,~~ \,\,\,\,\,\,
\label{bw215}
\\
A_{nr}(s)
 &\!\! =\!\! & N_{nr}\cdot (1-e^{-(\mu / a_1)^4}) 
                        \cdot (1 + a_2\mu + a_3\mu^2), 
\label{contin} 
\\
 \mu        &\!\! =\!\! & \sqrt{s} - m_0,~~ P(s) = \sqrt{1 - m_0^2/s } \nonumber
\end{eqnarray}
where
$N_{nr}$ normalizes the amplitude of the non-resonant spectrum,
$\sigma_0$ is a peak cross section for the hypothesized resonance,
and $m_x$, $\Gamma_x$ and $-\psi_x$ are the mass, total width and relative   
phase of the non-resonant amplitude to the standard Breit-Wigner 
amplitude.
The factor $P(s)$ gives a good approximation
of the two-body phase space factor for particles with similar masses;
both the $\phi(1020)$ and $f_0(980)$ have small but finite widths, and our selection
cut of $m(\pi\pi)\! >\! 0.85~\gevcc$ defines an effective minimum
mass, $m_0\! =\! 1.8$~\gevcc.
The form of 
$A_{nr}$ is determined from a simulation that takes the $\phi$ and 
$f_0(980)$ lineshapes into account. 
A very sharp exponential cutoff (parameter $a_1$) is needed to
describe the simulation well, but does not affect the spectrum 
well above threshold. 
There is no theoretical prediction for the form at high $s$,
other than that, in the absence of resonances, it should
fall smoothly with increasing $s$.
A second order polynominal (parameters $a_2$ and $a_3$) describes the
simulation, so
we fit Eq.~\ref{contin} to the data, floating $N_{nr}$, $a_1$, $a_2$ and $a_3$.
The result without a resonant component is shown as the dashed curve in Fig.~\ref{phif0xsall}.
The $\chisq_0=\! 80.5/(56-5)$ has confidence level $P(\chisq_0) = 0.0053$,
and the fitted parameter values are close to those from the
simulation;
it is unlikely that a simple, smooth threshold curve can accomodate
the data. 

Including a single resonance (Eq.~\ref{bw215}), we obtain a good fit with 
$\chisq_x=\! 37.6/(56-9)$ ($P(\chisq_x) = 0.84$), shown as the solid line
in Fig.~\ref{phif0xsall}.
The fitted resonance parameter values are
\begin{center}
$  \sigma_0\! =\!  0.13\pm 0.04\pm 0.02$~nb, \\
$     m_{x}\! =\! 2.175\pm 0.010\pm 0.015$~\gevcc , \\
$ \Gamma_{x}\! =\! 0.058\pm 0.016\pm 0.020$~\gevcc , and \\
$  \psi_{x}\! =\! - 0.57 \pm 0.30\pm 0.20$~rad. 
\end{center}
The first error is statistical and the second
is systematic.
Monte Carlo simulations show that the probability of
such a signal arising by chance is less than $10^{-3}$.
The modestly negative value of $\psi_x$ provides constructive interference
below the resonance peak and destructive interference above it, in 
accord with the data.
Variations in the resonance parameters are used to estimate the 
systematic errors.
The fit of the mass spectra in Fig.~\ref{phif01}a,b with
Eq.~\ref{bw215}  with normalization to the number of events
under the Breit-Wigner curve gives $170\pm63$ and $31\pm15$ events
for $\pipi$ and $\ppz$ respectively. 
Note that the observed structure is close to the $\Lambda\bar\Lambda$
production threshold at 2.23~\gevcc and the opening of this channel may
also contribute to the $\phi f_0$ cross section.

We perform a number of systematic checks.
Treating selected $\Kp\Km\Kp\Km$ and $\pipi\pi\pi$ events as signal,
we observe no structure.
Selecting $K^*(892)K\pi$ events, which have little kinematic overlap
with $\phi f_0(980)$, we see no structure.
Excluding the dominant $K^*(892)K\pi$ intermediate states and
selecting events with $m(\pipi)$ in the range 0.6--0.85~\gevcc
for the charged  mode we observe structure at 2.15~\gevcc  
with a similar yield. Because of the many overlapping 
intermediate states we cannot perform a quantitative measurement. 
This will be the subject of future investigation. 
Events with no $f_0(980)$ candidate do not exhibit a structure 
in the $\Kp\Km\ppz$ mode.
We conclude that the new structure decays to $\phi f_0(980)$ with
a relatively large branching fraction. We estimate 
$$ \BR_{x \to\phi f_0}\cdot\Gamma^{x}_{ee} =\frac{\Gamma_x \sigma_0
  m_x^2}{12\pi C } = (2.5\pm 0.8\pm 0.4)~\ev\ , $$
 where we fit the
product $\Gamma_x \sigma_0$ to reduce correlations, and 
the conversion constant $C=0.389\ \mathrm{mb\mbox{~}(\!\gevcc})^2$.

In summary, we present the most precise measurements of the cross sections 
for $\epem \!\to \Kp\Km\pipi$ and $\epem \!\to \Kp\Km\ppz$ 
from threshold to 4.5~\gev.
In the $\phi\pi\pi$ channels we observe the $J/\psi$ and $\psi(2S)$ 
but not the $Y(4260)$.
In the $\phi f_0$ channel, we observe a new resonance-like
structure, 
which might be interpreted as an \ssbar analogue of the $Y(4260)$, 
or as an $\ssbar\ssbar$ state that decays predominantly to $\phi f_0(980)$.

We are grateful for the excellent luminosity and machine conditions
provided by our \pep2\ colleagues, 
and for the substantial dedicated effort from
the computing organizations that support \babar.
The collaborating institutions wish to thank 
SLAC for its support and kind hospitality. 
This work is supported by
DOE
and NSF (USA),
NSERC (Canada),
IHEP (China),
CEA and
CNRS-IN2P3
(France),
BMBF and DFG
(Germany),
INFN (Italy),
FOM (The Netherlands),
NFR (Norway),
MIST (Russia),
MEC (Spain), and
PPARC (United Kingdom). 
Individuals have received support from the
Marie Curie EIF (European Union) and
the A.~P.~Sloan Foundation.


\vspace{-0.5cm}
\begin{thebibliography}{99}

\bibitem{y4260} \babar\ Collaboration, B.\ Aubert {\em et al.},
Phys. Rev. Lett. {\bf 95}, 142001 (2005). 

\bibitem{shin} Shi-Lin ~Zhu, Phys. Lett. {\bf B625}, 212 (2005). 

\bibitem{isr4pi}
\babar\ Collaboration, B.\ Aubert {\em et al.}, 
Phys. Rev. {\bf D71}, 052001 (2005).

\bibitem{babar} 
\babar\ Collaboration, B.\ Aubert {\em et al.}, 
Nucl.\ Instrum.\ Meth.\ {\bf A479}, 1 (2002).

\bibitem{kuehn2}
H.~Czy\.z and J.~H.~K\"uhn, Eur. Phys. J. {\bf C18}, 497 (2001).

\bibitem{kuraev} 
A.~B.~Arbuzov {\em et al.}, J. High Energy Phys. {\bf 9710}, 001 (1997).

\bibitem{strfun} 
M.~Caffo, H.~Czy\.z and E.~Remiddi, 
Nuovo Cim. {\bf A110}, 515  (1997); Phys. Lett. {\bf B327}, 369 (1994). 

\bibitem{PHOTOS} 
E.~Barberio, B.~van~Eijk and Z.~Was, 
Comput. Phys. Commun. {\bf 66}, 115 (1991).

\bibitem{GEANT4} 
GEANT4 Collaboration, S.\ Agostinelli {\em et al.},
\nima{506}, 250 (2003).

\bibitem{jetset} T.~Sj\"ostrand, Comput. Phys. Commun. {\bf 82}, 74 (1994).

\bibitem{koralb} 
S.~Jadach and Z.~Was, Comput. Phys. Commun. {\bf 85}, 453 (1995).

\bibitem{evtgen}
D.~J.~Lange, Nucl.\ Instrum.\ Meth.\ {\bf A462}, 152 (2001).

\bibitem{isr6pi} 
\babar\ Collaboration, B.\ Aubert {\em et al.},
Phys. Rev. {\bf D73}, 052003 (2006).

\bibitem{2k2pidm1} DM1 Collaboration, A. Cordier {\em et al.},   
Phys. Lett. {\bf B110}, 335 (1982). 

\bibitem{PDG}
Review of Particle Physics, S.\ Eidelman {\em et al.}, 
Phys. Lett. {\bf B592}, 1 (2004).

\bibitem{ycleo} CLEO Collaboration, T.\ E.\ Coan {\em et al.},
Phys. Rev. Lett. {\bf 96}, 162003 (2006). 

\end{thebibliography}
\end{document}